\documentclass[twocolumn,showpacs,amsmath,amssymb,prl,superscriptaddress, floatfix]{revtex4}

\usepackage{graphicx}
\usepackage{dcolumn}
\usepackage{bm}

\usepackage{amssymb}
\usepackage{amsmath}
\usepackage{amsfonts}
\usepackage{amssymb}
\usepackage{graphicx}
\usepackage{hyperref}
\usepackage{color}

\usepackage{xcolor}
\definecolor{red}{rgb}{0.7,0,0}
\definecolor{green}{rgb}{0.,0.35,0.}
\definecolor{blue}{rgb}{0.2,0.2,0.7} 
\definecolor{black}{rgb}{0.15,0.15,.15}

\begin{document}

\title{Quantum Phases of Cold Polar Molecules in 2D Optical Lattices}

\author{B. Capogrosso-Sansone} 
\affiliation{ITAMP, Harvard-Smithsonian Center of Astrophysics, Cambridge, MA, 02138, USA}
\author{C. Trefzger}
\affiliation{ICREA and ICFO - Institut de Ciencies Fotoniques, 08860 Castelldefels (Barcelona), Spain}
\author{M. Lewenstein}
\affiliation{ICREA and ICFO - Institut de Ciencies Fotoniques, 08860 Castelldefels (Barcelona), Spain}
\author{P. Zoller}
\affiliation{IQOQI and Institute for
Theoretical Physics, University of Innsbruck, 6020 Innsbruck, Austria}
\author{G. Pupillo}
\affiliation{IQOQI and Institute for
Theoretical Physics, University of Innsbruck, 6020 Innsbruck, Austria}
\begin{abstract}
We study the quantum phases of hard-core bosonic polar molecules on a two-dimensional
square lattice interacting via repulsive dipole-dipole interactions. In the limit of
small tunneling, we find evidence for a devil's staircase, where Mott-solids appear at
rational fillings of the lattice. For finite tunneling, we establish the existence of
extended regions of parameters where the groundstate is a supersolid, obtained by doping
the solids either with particles or vacancies. We discuss effects of finite temperature and finite-size confining potentials as relevant to experiments. 
\end{abstract}
\parindent  8.mm
\maketitle

Trapped atomic and molecular quantum gases allow the realization of quantum lattice
models of strongly interacting particles~\cite{BlochReview2005}.
The preparation of quantum gases of groundstate polar molecules with strong electric dipole
moments opens the way to the study of lattice models with tunable long-range interactions
in atomic-molecular-optical (AMO) setups~\cite{PolMolExp,LewensteinReview}. In this context, the challenges are to (i)
identify the fundamental Hamiltonians underlying the physics of strongly interacting dipolar gases,
(ii) analyze the associated quantum phases in the context of their realization in AMO setups.
This entails (iii) studying the preparation and detection of these phases in the presence of
finite-size confining potentials and finite temperature, as relevant to experiments.

In this letter we analyze the microscopic tight-binding Hamiltonian realizable
with polar molecules trapped in optical lattices under collisional stability.
This is a 2D Hubbard-like model for hard-core particles with infinite-range interactions which
we show to display novel quantum phenomena with no counterpart in the atomic case.
These include: (a) a devil's staircase (DS) of Mott solids at rational lattice fillings,
and (b) supersolid phases (SS) obtained by doping solids with either vacancies or additional particles.
The infinite-range interactions stabilize the SS as the low-energy phase for a
large range of system's parameters, raising interesting prospects for its realization
with polar molecules. While various kinds of SS phases (but not the DS) have been found
in models with shorter-range interactions~\cite{SScite,Batrouni1995,Comment00}
[e.g., nearest-neighbor (NN) interactions  with soft-core particles, or NNN interactions],
here the emphasis is in simulating experimentally relevant systems with up to $N\sim 10^3$ particles,
at finite temperature, and for finite-sizes with harmonic confinement.
Results for experimental observables with current AMO technology are presented.

\begin{figure}[b!]
\vspace*{-0.5cm}
\center{\includegraphics[width=0.75\columnwidth]{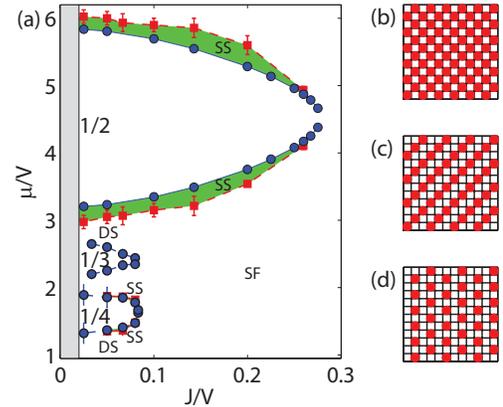}}
\caption{(color online) (a) Phase diagram of model~\eqref{H} with $\Omega=0$ and at $T=0$. 
Lobes: Mott solids (densities indicated); SS: supersolid phase; SF:
superfluid phase. DS: devil's staircase.
Panels (b-d): sketches of the groundstate configuration for the Mott solids in
panel (a), with density $\rho=1/2, 1/3$ and 1/4, respectively. }\label{fig_lobes}
\end{figure}

We consider $N$ bosonic polar molecules aligned by a static electric field with induced dipole moment $d=\sqrt{D}$, implying strong dipole-dipole interactions. Collisional stability is reached confining the molecules to a 2D plane, using a strong transverse trapping field, e.g a 1D optical lattice, with a harmonic oscillator frequency $\omega_{\perp}$ to prevent collapse due to attractive forces between aligned dipoles, for interparticle distances larger than $ a_{\rm min}=(12 D/m \omega_\perp^2)^{1/5}$, with $m$ the mass of a molecule~\cite{Micheli07}.
An additional 2D optical lattice with spacing $a> a_{\rm min}$ and harmonic oscillator frequency $\omega$ confines the particles in-plane. The following single-band Hamiltonian for hard-core bosons on a 2D square lattice is then obtained provided $\hbar \omega_{\perp} \gg \hbar \omega  > \{D/a^3 , k_B T$\}, with $T$ the temperature, and with the requirement that the initial system has no doubly
occupied sites~\cite{BuechlerNatPhys2007}
\begin{equation}
 H = -J \sum_{<i,j>} b^{\dag}_{i} b^{}_{j}
+ V\sum_{i<j}\frac{n_{i} n_{j}}{r^3_{ij}}  - \sum_i (\mu-\Omega \; i^2) n_i\;. 
\label{H}
\end{equation}
The first and second terms in Eq.~\eqref{H} describe the standard kinetic energy with hopping
rate $J$ and the repulsive dipole-dipole interaction with strength $V=D/a^3$ and $r_{ij}=|i-j|$, respectively; $b_{i}$ and $b^{\dag}_{i}$ are bosonic operators with $b^{\dag 2}_{i}=0$ and $n_{i}
= b^{\dag}_{i} b_{i}$; 
$\mu$ is the chemical potential and $\Omega=m\omega^2a^2/2$. For a gas of RbCs polar molecules ($d=1.25$Debye) with transverse and in-plane trapping $V_{0,\perp}/E_{\rm R}=40$ and $V_{0}/E_{\rm R}=4$, respectively, and lattice spacing $a=400$nm $> a_{\rm min}\sim 100$nm, $\omega_{\perp}/2 \pi \gg \omega/2 \pi > D/(a^3 h)$, with $D/(a^3 h)\simeq 3.5$kHz, and thus the single-band Hamiltonian Eq.~\eqref{H} is valid, provided $T \lesssim 200$nK. With an in-plane tunneling rate $J/h\simeq 120$Hz, the ratio $J/V$ can be tuned to any value $J/V \gtrsim 0.03$ by varying the strength of the DC field. We have studied the quantum phases of Eq.~\eqref{H} by means of large scale Monte-Carlo simulations based on the Worm Algorithm~\cite{Worm}, using no cutoff in the range of the dipole-dipole interaction~\cite{Comment0}.
While the focus is on experimentally-relevant trapped systems with finite $T$, we find it convenient to first discuss the phase-diagram in the homogeneous situation, and then use these results to explain the phases for $\Omega \neq 0$, and observables in experiments.
This is followed by more detailed discussions of some aspects of the various phases.

{\em Homogeneous case:} Our exact results for $\Omega=0$ are summarized in
Fig.~\ref{fig_lobes}(a), the computed $T=0$ phase-diagram as a
function of $\mu$ and $J$, with $1<\mu/V<6$ and $J/V > 0.02$ (unshaded
area). For small-enough hopping $J/V \ll 0.1$, we find that the low-energy phase
is incompressible ($\partial \rho / \partial \mu = 0$, with $\rho$ the filling-factor)
for most values of $\mu$.
This parameter region is labeled as DS in the figure and corresponds to the
classical {\it devil's staircase}, i.e. a succession of incompressible ground
states, dense in the interval $0 < \rho <1$, with a spatial structure
commensurate with the lattice for all rational fillings~\cite{devil's}, with no
analogue for shorter-range interactions. For finite $J$, three main solid Mott
lobes emerge with $\rho=1/2$, 1/3, and 1/4, named checkerboard (CB), stripe, and star solids, respectively. The corresponding groundstate configurations are sketched in panels (b-d). These large Mott lobes are found to
be robust in the presence of a confining potential and finite $T$ (see below),
and thus are relevant to experiments. We notice that the shape of the 
solids with $\rho=1/2$ and 1/4 away from the tip of the lobe can be shown to be qualitatively captured
by mean-field calculations, while this is not the case for the stripe solid at
filling 1/3 which has a pointy-like structure characteristic of
fluctuation-dominated 1D configurations. Mott lobes at other rational filling
factors, e.g. $\rho=1$ and 7/24, are present, however not shown here.

\begin{figure}[b]
\vspace*{-0.6cm}\begin{center}
\includegraphics[width=0.8\columnwidth]{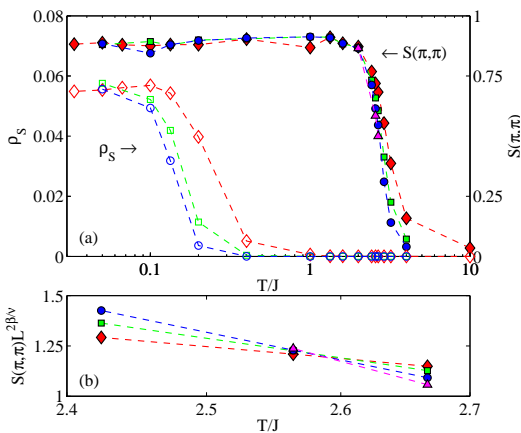}
\end{center}
\vspace{-0.5cm}\caption{(color online) Finite-$T$ melting of SS at $J/V=0.1$: (a) $\rho_{\rm s}$ (empty symbols) and  $S(\pi,\pi)$ (full symbols) vs. $\rho$, for $L=8,12, 16$ and 20 (diamonds, squares, dots, and triangles, respectively); (b) $S(\textbf{k})L^{2 \beta/\nu}$ vs. $\rho$, with $2 \beta/\nu=1/4$. }\label{fig_SS_finiteT_V10}
\end{figure}

For large enough $J/V$, the low-energy phase is superfluid, for all $\mu$.
At intermediate values of $J/V$, however, we find that by doping the Mott solids
{\it either with vacancies} (removing particles) {\it or interstitials} (adding extra particles)
a supersolid phase can be stabilized, with coexisting superfluid and crystalline orders
(we find no evidence of SS in the absence of doping).
We find that the solid/superfluid transition consists of two steps, with both transitions of the second-order and the supersolid as the intermediate phase
(see Fig.~\ref{fig_SSV5} below). Remarkably, the long-range interactions stabilize
the supersolid in a wide range of parameters. For example, a vacancy SS is present
 for densities $0.5> \rho \gtrsim 0.43$, roughly independent of the interaction strength.
 For experiments, a fundamental question is the observability of the phases described
above for finite $T$. In particular, for the SS phase we show in Fig.~\ref{fig_SS_finiteT_V10}
 that by increasing $T$ it melts into a featureless normal fluid via a two-step transition,
 the intermediate phase being a normal fluid with finite density modulations, similar to a liquid crystal.

\begin{figure}
\includegraphics[width=0.7\columnwidth]{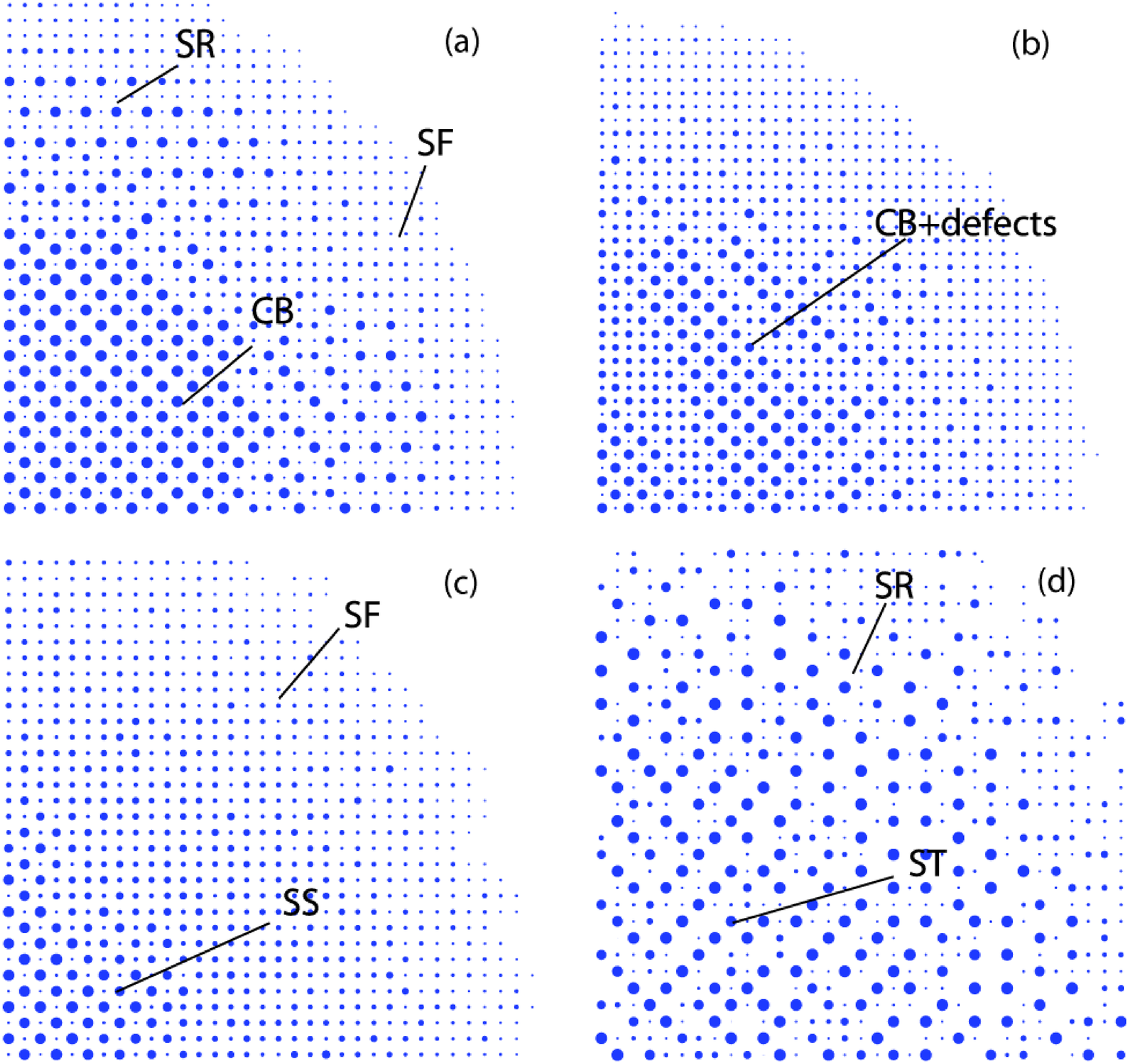}
\caption{(color online) Spatial density profile in 2D for $N\simeq 1000$ particles
 in a harmonic potential. Phases are indicated (CB, SR, ST stand for checkerboard, star, and stripe solids respectively). (a-b) $V/J=15$, $\mu/J=55$,
$\Omega/J=0.05$ and $T/J=0.0377$, with temperature annealing performed in panel a); (c) $V/J=5$, $\mu/J=19$, $\Omega/J=0.01$ and $T/J=0.1$;
(d) $V/J=20$, $\mu/J=51$, $\Omega/J=0.04$ and $T/J=0.25$. }\label{trap}
\end{figure}
{\em Harmonic trap and experimental observables:} The recent experimental achievement
of single-site  addressability in optical lattices using electron and optical microscopy
allows for a direct, {\em in-situ}, observation of particle positions and particle-particle
correlations in experiments~\cite{Ott2009}. Thus, the key observables for present and future
experiments are the {\em in-situ} density distribution and particle correlations, from which
the phases above can be detected. The question is how the phases described in Fig.~\ref{fig_lobes}
will be seen in an experiment. Here, we provide snapshots of particle configurations for realistic
experimental situations with $N\sim 10^3$ particles trapped with harmonic confinement, and small finite $T$.

In Fig.~\ref{trap} we show snapshots of the spatial density distribution in the
lattice (shown is a single quadrant). Each circle corresponds to a different
site, and its radius is proportional to the local density. In panels a) and b),
$\mu$ has been chosen such that particles at the trap center are in the CB
phase, with very small $T$. The density profile shows a wedding-cake structure,
with concentric Mott-lobes with density $\rho=1/2$ and 1/4, analogous to the
shells with contact interactions~\cite{Batrouni2002}. However, while the system
parameters are the same in both figures, panel (a) shows regular CB and star
patterns, while in panel (b) extended defects are present in the CB phase and
the star is barely visible. This is due to the different preparation of the states
in panels (a) and (b). In fact, in panel (a) we performed temperature annealing
of the system prior to taking the snapshot, while this was not done in panel
(b). We find that the defects in (b) reflect the existence of a large number of
low-energy metastable states, which are a direct consequence of the long-range
nature of the interactions, and will be of relevance for experiments.

Supersolid and stripe phases are shown in panels (c) and (d), respectively. In panel (c), $\mu$ has been chosen to realize an extended vacancy-SS region, surrounded by a superfluid. We notice that here a finite $T=0.1J$ has been chosen, compatible with the existence of the SS phase (see below). The density-distribution in the vacancy-SS looks similar to the {\it ordered} CB phase, even without annealing. Self-annealing is in fact here enabled by the (small) superfluid component of the SS phase. Small coherence peaks will be present in time-of-flight experiments, allowing for a clear determination of this phase.
Panel (d) shows a disordered stripe-phase at the center, surrounded by an extended Mott-shell with $\rho=1/4$. The disorder in this case is a result of both finite $T/J=0.25$ and the fact that the stripe solid is less robust towards quantum and classical fluctuations compared to the CB and star ones.

These exact results for $\Omega \neq 0$ confirm that the phase-diagram Fig.~\ref{fig_lobes} is key to predict and interpret experimental observables, assuming a local density approximation. In the remainder of this work, we provide more details on the phases described above.

\begin{figure}[t!]
\center{\includegraphics[width=0.8\columnwidth]{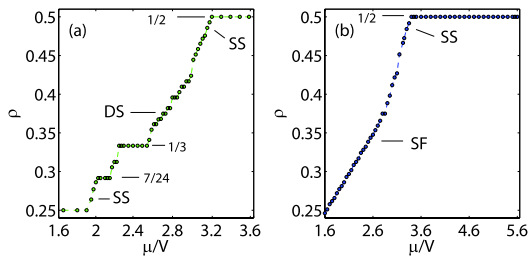}}
\vspace{-0.3cm}\caption{(color online) $\rho$ vs. $\mu$.
(a):  Solids and SS for a system with linear size $L=12$ and $J/V=0.05$. Some $\rho$ are indicated. (b): superfluid and vacancy-SS for $L=16$ and $J/V=0.1$.}
\label{fig_n_vs_muL12V20}
\end{figure}

{\it Incompressible phases:} For each Mott lobe in Fig.~\ref{fig_lobes}(a), the
solid order is characterized by a finite value of the structure factor $
S(\textbf{k})=\sum_{\textbf{r},\textbf{r}'}
\exp\left[i\textbf{k}\left(\textbf{r}-\textbf{r}'\right)\right]\langle
n_{\textbf{r}} n_{\textbf{r}'}\rangle/N$, with \textbf{k} the reciprocal lattice
vector for each solid. For the CB, stripe, star solids, this is $(\pi,\pi)$,
$(\pm2\pi/3,2\pi/3)$, and $(\pi,0)$ [or $(0,\pi)$], respectively. The boundaries
of the Mott lobes have been calculated from the zero momentum Green function
(see e.g.~\cite{GreenFunction}), for linear system sizes up to $L=20$ (CB and star
lobes), and from $\rho(\mu)$-curves with sizes up to $L=24$ (stripe lobe). 

Interestingly, we find evidence for incompressible phases in addition to those
corresponding to the lobes in Fig.~\ref{fig_lobes}. This is shown in
Fig.~\ref{fig_n_vs_muL12V20}(a,b), where the particle density $\rho$ is plotted as a
function of the chemical potential $\mu$ for  $J/V=0.05$, and 0.1, respectively. In the figure, a continuous increase of $\rho$ as a
function of $\mu$ signals a compressible phase, while a solid phase is
characterized by a constant $\rho$. 
The main plateaux in panel (a) correspond to the Mott lobes of Fig.~\ref{fig_lobes}, while the other steps are incompressible phases, with a fixed, {\it integer}, number of particles. This progression of steps is an indication of a devil's-like staircase in the density, the latter being fully realized in the classical limit of zero hopping.

Since the simulations are necessarily performed for finite $L$ (with periodic boundary conditions) and $T$, only the lobes with comparatively short periodicity (e.g., $\rho=1/2$ and 1/4) and sizeable gaps will be resolved in the calculations. Consistently, we find that determining the groundstate configuration for each DS-step directly from the simulation is often challenging, since: {\it i)} for many rational fillings [e.g. $\rho=7/24$ in panel (a)] it would require to consider system sizes (much) larger than those considered here, and {\it ii)} the long-range interactions determine the presence of numerous low-energy \textit{metastable} states~\cite{Menotti2007}, which for finite $T$ can result in the presence of defects or in disordered structures. 
However, we note that the practical relevance of possible Mott lobes with large periodicity in the DS region is somewhat limited, since they will most likely {\it not} be observable in experiments with trapped molecules, as shown above.

{\it Supersolid phases:} A very different situation is shown by the behavior of $\rho(\mu)$ immediately above the star plateau of panel (a), and below the CB plateau of panel (b) in Fig.~\ref{fig_n_vs_muL12V20}. Here the density grows smoothly with increasing $\mu$, signaling a fluid phase, and we find that the superfluid stiffness $\rho_{\rm{s}}= T \langle W^2 \rangle$ is finite, with $W$ the winding number. Remarkably, and in contrast to previous studies with shorter-range interactions, we find that the appropriate structure factor $S(\bf{k})$ corresponding to these solids remains finite for chemical potentials $\mu$ below and above each Mott lobe, signaling the existence of both vacancy-induced and particle-induced supersolidity, with no indication of phase-separation. We measured supersolid behavior around both Mott lobes with $\rho=1/2$ and 1/4, for $J/V \gtrsim 0.05$ and 0.067, respectively.
As an example, Fig.~\ref{fig_SSV5}(a) shows the case of vacancy-induced supersolidity for $J/V=0.2$.
For an extended range of densities, both the superfluid stiffness $\rho_s$ and the static structure factor $S(\pi,\pi)$, are finite and size-independent. We find evidence that the SS melts into a superfluid via a second-order Ising-type quantum phase-transition. This is shown by the crossing of the curves in panel (b) of Fig.~\ref{fig_SSV5}, where we plot $S(\textbf{k})L^{2 \beta/\nu}$ as a function of density (the critical exponents  $2\beta/\nu=1.0366(8)$ correspond to the three-dimensional Ising universality class~\cite{Martin}).
The supersolid behavior persists for smaller $J/V$-ratios, however $\rho_{\rm s}$ tends to decrease with $J/V$. While the results above point to a generic mechanism for solid/liquid transitions in 2D with the SS as the intermediate phase~\cite{Spivak}, we have not found evidence of SS phases below and above the stripe lobe. This may be due to an extremely low $T$ for superfluidity here (we have used $T_{\rm min}=J/3L$), or to the strong 1D character of the stripe lobe.
\begin{figure}[t!]
\vspace*{-0.30cm}
\center{\includegraphics[width=0.8 \columnwidth]{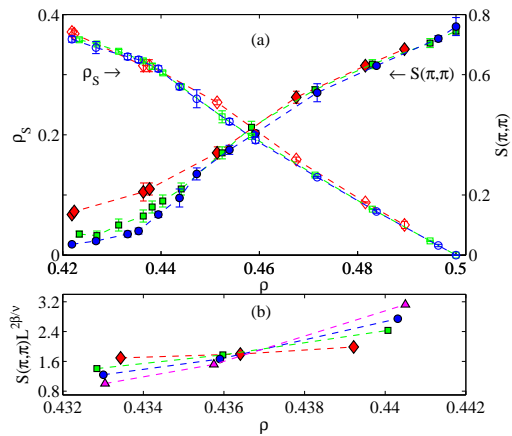}}
\vspace{-0.4cm}\caption{(color online) Vacancy supersolid for $J/V=0.2$: (a) $\rho_{\rm s}$ (empty symbols) and $S(\pi,\pi)$ (full symbols) vs. $\rho$, for $L=8,12,16$ and 20 (diamonds, squares, dots, and triangles, respectively); (b) $S(\textbf{k})L^{2 \beta/\nu}$ vs. $\rho$, with $2 \beta/\nu=1.0366$. The crossing indicates an Ising-type second-order transition.}\label{fig_SSV5}
\end{figure}

{\it Finite-$T$:} We studied the melting of the SS into a normal phase with increasing $T$, for the case of vacancy supersolidity below the CB solid, with $J/V=0.1$. Figure~\ref{fig_SS_finiteT_V10} shows  $\rho_{\rm s}$ and $S(\pi,\pi)$ vs $T$. The melting of the SS proceeds through two successive transitions. First, SS melts into a liquid-like phase reminiscent of a liquid crystal, with zero $\rho_{\rm s}$ and finite $S(\pi,\pi)$. The drop of $\rho_{\rm s}$ for $T\simeq 0.1 J$ in Fig.~\ref{fig_SS_finiteT_V10} signals a transition of the Kosterlitz-Thouless type, with critical temperature $T_{\rm KT}=\pi \rho_{\rm s} \hbar^2 \rho /2 m$, and $m=1/2Ja^2$. Upon further increasing temperature, $S(\pi,\pi)$ drops to zero for $T\simeq 2.6J$. In panel (b) we show that this is consistent with an Ising-type transition, by plotting the expected scaling for $S(\pi,\pi)$ in two dimensions (here, $2\beta/\nu=1/4$).

In conclusion, we have shown that polar molecules on a square lattice will realize exotic solid and supersolid quantum phases under realistic experimental conditions.

{\it Note added:} While completing the present work, we became aware of a simultaneous, independent study on a {\it triangular} lattice with $\Omega=0$, see Ref.~\cite{Pollet2009}.

We thank N. Prokof'ev and the authors of Ref.~\cite{Pollet2009} for fruitful discussions. This work was supported by ITAMP, MURI, EOARD, U. Md. PFC/JQI, the Austrian FWF, the EU (STREP FP7-ICT-2007-C, NAME-QUAM), and the Spanish MEC (FIS2008-00784, QOIT).


\begin{thebibliography}{50}
\bibitem{Comment0}An Ewald summation is used in the interaction potential.
\end{thebibliography}
\end{document}